# Basic considerations in the design of an electrostatic electron monochromator

M.J. Adriaans[a,*], J.P. Hoogenboom[a], and A. Mohammadi-Gheidari[a]

[a]Department of Imaging Physics, Delft University of Technology, Delft, The Netherlands

[*]Corresponding author, E-mail address: m.adriaans@tudelft.nl

## Abstract

Monochromators are an essential component in electron microscopy and spectroscopy for enhancing the spatial and energy resolution. However, its adoption in scanning electron microscopes (SEMs) remains limited because of its high cost and operational complexity. Through a thin-deflector analysis of an electrostatic homogeneous-field deflector, the extreme sensitivity of current monochromators to power supply drift and mechanical imperfections is demonstrated. These stringent alignment requirements for achieving optimal energy resolution often necessitate the use of additional correcting elements, adding to both cost and complexity. We demonstrate that the fringe-field deflector is instead less sensitive to these issues. Hence, a cost effective and simple monochromator design approach based on pure fringe fields is proposed. This monochromator doesn't need extra correcting elements and its optimal energy resolution is achieved by including momentary deceleration lenses surrounding the main deflector. This fully electrostatic design could be realized using MEMS technology, offering a simpler and more accessible approach for filtering beam energies.

# 1 Introduction

In recent years, there has been growing interest in Low-Voltage-Electron-Microscopy (LVEM), particularly Low-Voltage-Scanning-Electron-Microscopy (LVSEM), for applications such as imaging the surface of charging sample [1], [2]. However, as the beam energy decreases, chromatic aberration blur significantly degrades the resolution, especially at extremely low landing energies, in the range of only a few hundred electron-volts (eV). Reducing the energy spread of the electron source greatly mitigates this issue, as illustrated in Figure 1a. The figure shows the variation of the axial FW50 probe diameter, $d_p$, which contains 50% of the total probe current as a function of the beam's opening angle, $\alpha$. To calculate $d_p$ different contributions are added according to [3]

$$d_p = \left( \left[ d_{geo}^{1.3} + (d_\lambda^4 + d_s^4)^{\frac{1.3}{4}} \right]^{\frac{2}{1.3}} + d_c^2 \right)^{\frac{1}{2}}, \qquad 1$$

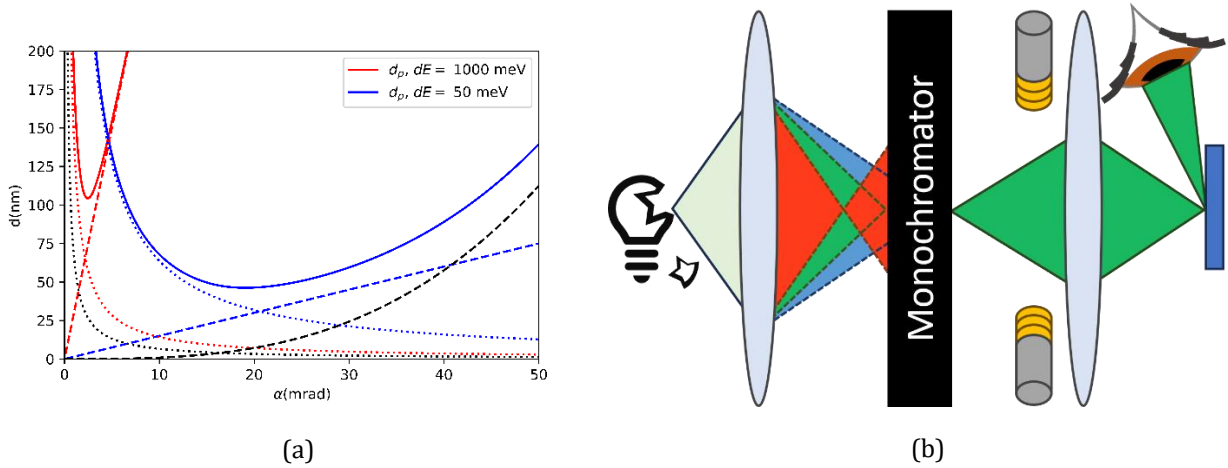

(a)            (b)

*Figure 1: (a) FW50 probe size for different dE (red for 1000 meV and blue for 50 meV) due to a combination of geometric spot size (decreasing dotted red and blue curves) and chromatic aberration (increasing dashed red and blue curves), diffraction (decreasing dotted black curve) and spherical aberration (increasing dashed black curve) leading to a combined total spot size indicated by the unbroken red and blue lines. Here we take $C_s = C_c = 5mm$, (fixed) probe current of $0.5\ nA$ and $E = 100\ eV$. (b) Schematic illustration of a monochromated SEM column. The source (depicted as a broken light bulb) emits an unfiltered electron beam and in a first crossover the spot is blurred due to the chromatic aberration of the first illustrated lens. A monochromator (displayed as a black box module) removes the electrons with higher (blue) and lower (red) kinetic energies than the nominal (green) electron beam energy. The filtered beam is then focused onto and scanned across a sample plane, where secondary electrons are created and then detected by an observer (eye).*

where $d_{geo} = \frac{2}{\pi}\sqrt{\frac{I}{B_r \phi}}\frac{1}{\alpha}$ is the FW50 size of the geometric source image, $d_c = 0.6 C_c \frac{dE}{E} \alpha$ is the FW50 size of the chromatic aberration blur, $d_s = 0.18 C_s \alpha^3$ is the FW50 size of the spherical aberration blur and $d_\lambda = 0.54 \frac{\lambda}{\alpha}$ is the FW50 size of the diffraction blur. In these expressions, $B_r$ is the reduced brightness of the electron beam, $\phi$ is the acceleration potential ($E = e\phi$ = acceleration energy), $dE$ is the energy spread of the electron source, $\lambda$ is the wavelength of the electrons, $C_s$ and $C_c$ are the spherical and chromatic aberration coefficients of the objective lens respectively.

As demonstrated in Figure 1a, reducing $dE$, e.g. through incorporating a monochromator, (as shown schematically shown in Figure 1b) improves the spatial resolution of LVSEM's. Monochromators are widely used in High-Resolution-Electron-Energy-Loss-Spectroscopy (EELS) to improve energy resolution [1] and in (Scanning) Transmission Electron Microscope ((S)TEM) to improve spatial resolution [2]. However, this isn't particularly true for SEM's and the main reason for it is the higher cost and complexity of current monochromator designs. There are various monochromator designs specifically used in (S)TEM and EELS[4] known as Alpha-, Omega- and Wien-type monochromator as shown schematically in Figure 2. These different monochromator layouts have previously been described and compared[5]. Each configuration begins with an unfiltered beam emitted from a (virtual) source, which is then collimated by a lens. The lenses, play a crucial role in converting angular dispersion into positional dispersion at the aperture selection or slit plane, letting only the nominal energy to pass through while stopping the lower and higher energies, depicted in red and blue. Essentially, in the heart of all these monochromators there is a "uniform" magnetostatic or electrostatic or a combination of both deflection fields that creates angular dispersion of the beam. The performance of a monochromator is typically evaluated in terms of its theoretical energy resolution. Although that seems a natural number to describe monochromator performance, it is a misleading number to consider alone. One should also consider the costs and complexities of such systems in order to achieve such resolutions (e.g.

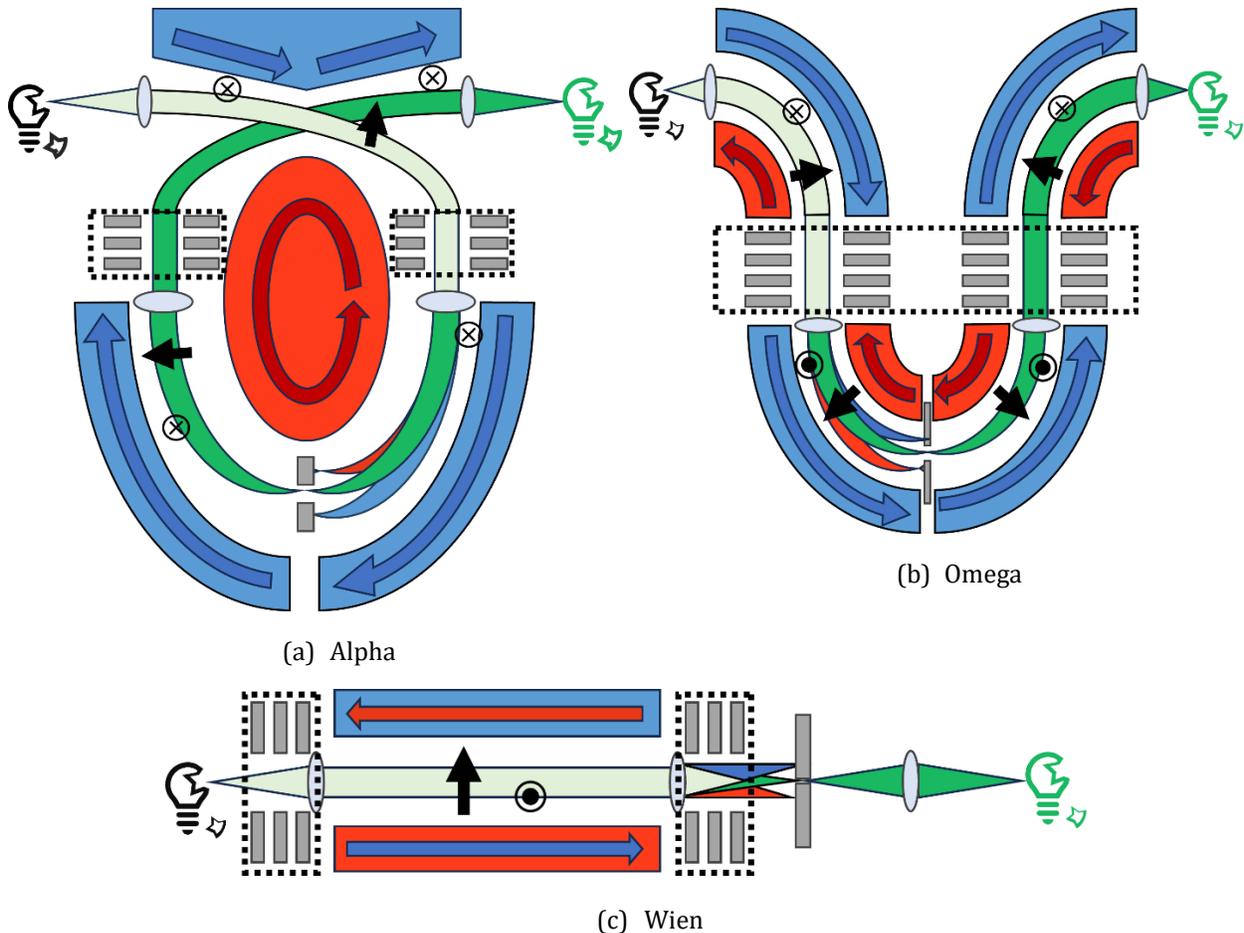

*Figure 2: Schematic representation of common monochromator layouts. Figure 2a shows an Alpha type, Figure 2b an Omega type, and Figure 2c a Wien type. The Alpha and Omega types are based on either magnetostatic or electrostatic fields. For the magnetostatic configuration, red and blue arrows show the direction of current through surrounding shapes, creating a magnetic field. In the electrostatic case, a voltage difference is applied, inducing a similar beam curvature. In the Wien type, opposing magnetostatic and electrostatic fields create dispersion where electrons with nominal beam energy remain undeflected.*

energy resolution × price). In practice, the monochromators are much more complicated than those deflection units. For example, in these schematic monochromator depictions, the gray-dotted area indicates extra multipole correction elements necessary to compensate the undesired effects of mechanical imperfections and geometric aberrations. Almost all monochromators consist of such a set of complicated and expensive correction elements. To a great extent, the cost and complexity of monochromators depend on the level of complexity of correction elements added to the original theoretical design.

As mentioned earlier, due to higher cost and complexity of the current monochromators, they are not typically used in SEM columns. To the best of our knowledge, there is only one commercially available monochromator concept for SEM known as UC [7]. Unlike the former concepts, though with a limited energy resolution of around 60meV, UC is based on a relatively simple gun modification. For LVSEM, a better monochromator in terms of better energy resolution and lower cost and complexity is required.

It is the aim of this paper to explore why and how these monochromators are so expensive and complex. In sections 2 and 3 the main parameters influencing the dispersion resolution of the dispersive element, a simple electrostatic deflector, is analysed and their boundaries are discussed. In section 4, the basic layouts of future simple monochromator, free from all these issues, dedicated for LVSEM is proposed.

# 2 Requirements for an electrostatic deflector monochromator

In any monochromator, a (semi)collimated beam traverses a deflection field, acting as dispersive element, to create angular dispersion throughout the deflection field. This angular dispersion is then imaged onto a selection aperture or slit plane. Here, a simplified version of a monochromator based on a single electrostatic deflector will be discussed. Such a simplified layout is schematically depicted in Figure 3. Though the figure depicts a straight axis deflector, the final result is, to some extent, applicable to other forms, such as a curved axis deflector.

In a deflector, the deflection angle $\theta$, is a function of acceleration potential $\phi$. A kinetic energy spread $dE$, which we write from here on in terms of $d\phi = \frac{dE}{e}$ (with elementary charge $e$), causes a spread in the deflection angle $d\theta$ due to transverse chromatic aberration. Linear expansion of transverse electrostatic deflection angles around $\phi$ yields

$$\frac{d\phi}{\phi} = -\frac{d\theta}{\theta}. \qquad 2$$

This angular spread of different energies passing through a deflector is the main characteristic of deflector based monochromators. In order select an energy window, a lens at the exit plane of the deflector converts this angular dispersion to a spatial dispersion.
In the dispersion plane, a slit selects the desired energy window and cuts away the rest of the energy spectrum.
An ideal monochromator should introduce a large dispersion to allow for a small selectable energy window. In order to optimize for energy resolution, the size of the slit is designed to be the size of the focused probe of electrons with nominal beam energy. Ideally, the size of the probe at the slit plane should be equal to the geometric image of the source at that plane. However, this is not generally the case. The probe is larger than the geometric image of the source due to additional contributions, which degrades the energy resolution of the monochromator. Here we investigate the main parameters affecting the energy resolution of a monochromator. To do this, we only focus on the parameters influencing the performance of the main dispersion element, the deflector. Furthermore, we aim to limit our calculations to the case where no additional effort is spent to correct the limitations associated with deflector based monochromators as presented here. To isolate the contributions of the main deflector only, we study the impact of various parameters affecting the smallest discernible angle $d\theta$ at the deflector exit plane.

An analytical description of thin deflectors and a contribution to the smallest discernible angle due to geometric aberrations is presented in subsection 2.1. The same results are used to derive contributions of mechanical misalignments in subsection 2.2. Then, we derive contributions of electric potential supply instability in subsection 2.3 and discuss Boersch effect in subsection 2.4.

## 2.1 Geometric aberrations

The deflection of a collimated electron beam in a thin deflector is schematically shown in Figure 3. The collimated electron beam enters and leaves a deflector at an angle $\theta/2$ maintaining a constant $x$ coordinate throughout the deflection region[1]. The beam internal angle $\alpha$ is the ratio between the effective focal length of the collimating lens and the (virtual) source radius $r_s$. The radius of the beam $r$ is constant throughout the deflector. The deflection space is modelled as a homogeneous deflection field, with a sharp cutoff. This choice implies that the electrons are accelerated in the $z$-direction only upon entering and leaving the deflector. The effect of the fringe field in the $x$-direction is small compared to the deflection caused by the homogeneous field for narrow deflectors with weak excitation. Therefore, the effect of the shape of fringe fields on the monochromator resolution is ignored.

The change in (non-relativistic) momentum, $m\vec{v}$, of an electron with velocity $v$ through the Lorentz force is

$$\frac{1}{m}\frac{d\vec{p}}{dt} = \frac{d\vec{v}}{ds}v = \frac{e}{m}\vec{\nabla}\phi, \qquad 3$$

with $\phi$ the electrostatic potential, $e$ the elementary charge, $m$ the electron (rest-)mass and $ds$ an infinitesimal element along the trajectory of an electron. Fixing the ground potential as $e\phi = 1/2mv^2$ and integrating over a known particle trajectory $s$ yields

$$d\vec{v} = \frac{e}{m}\frac{1}{v}\vec{\nabla}\phi ds = \frac{1}{2}\eta^2 \frac{1}{\eta\phi^{\frac{1}{2}}}\vec{\nabla}\phi ds = \frac{\eta}{2}\vec{\nabla}\sqrt{\phi}ds \qquad 4$$

$$\Delta\vec{v} = \eta \int \vec{\nabla}\phi^{\frac{1}{2}} ds, \qquad 5$$

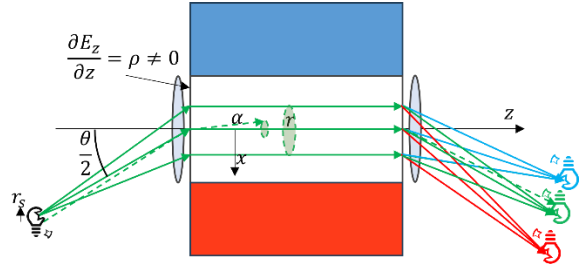

Figure 3: Schematic representation of the thin deflector approximation. The beam comes in at an angle, the transverse x-position of the particle is then assumed constant throughout the deflector, and the particle leaves the deflector with the same initial angle in the opposite direction. The lenses are added for demonstration of the positional dispersion. These lenses would in practice also add a chromatic defocus, illustrated here with different focal planes for the red and blue rays. The dashed green line indicates beam angle α is equal to the ratio between (virtual) source radius and effective focal length $r_s/f$ respectively. The effect of this angle is exagerrated in the schematic for visibility. The radius of the collimated beam is $r$.

---

[1] Typically, the constant $\phi$ before the deflector is equal to the constant $\phi$ after the deflector. Integration over a straight axis $x = x_0$ always yields $\Delta v_z = 0$ for systems symmetric in the middle. This causes the necessity $v_{x,0} = -v_{x,1}$ (before and after deflector). This, in turn, implies that a blurred spot in focus behind the deflector is caused by an equally blurred incoming beam before the deflector. A more general approach is presented in subsection A.1. Such procedures can improve accuracy of the calculation for large angles or include position aberrations in addition to the angle aberrations in the focal plane of the monochromator derived here, but currently leave the general arguments made about deflectors here unchanged.

[2] with[3] $\eta = \sqrt{\frac{2e}{m}}$.

For a homogeneous field deflector, the exact trajectory $s$ is a parabola and we can expand around the initial coordinates to retrieve aberrations as done before[7]. In this formulation, we can substitute any guess trajectory, including a parabola, to retrieve an approximate result. In order to keep this approach simple for generalized forms of $\phi$, we substitute $ds \approx dz$. This assumes the particle will travel along a constant $x$ (thin deflector approximation) through the deflector, and the trajectory is schematically depicted in Figure 3. Dividing $\Delta v_x$ by the velocity far away from the deflector $v_0 = \eta\sqrt{\phi_0}$, where the potential is a constant $\phi_0$, to yield the deflection angle $\theta \approx 2\sin\left(\frac{\theta}{2}\right) = \frac{\Delta v_x}{v_0}$ (small angle approximation), we get

$$\theta = \frac{1}{\sqrt{\phi_0}} \int \frac{\partial}{\partial x} \phi^{\frac{1}{2}} \, dz. \qquad 6$$

This then allows Taylor expansions of $x$-component of the integrand along $x$ as

$$\theta = \frac{1}{\sqrt{\phi_0}} \int dz \left[ O(x^3) + \frac{\phi'(0)}{2\phi(0)^{\frac{1}{2}}} + x \frac{2\phi(0)\phi(0) - \phi'(0)^2}{4\phi(0)^{\frac{3}{2}}} + x^2 \frac{4\phi(0)^2 \phi'''(0) + 3\phi'(0)^3 - 6\phi(0)\phi'(0)\phi''(0)}{16\phi(0)^{\frac{5}{2}}} \right] \qquad 7$$

In this expansion, the first term, the term with $x^0$, represents the deflection effect of the deflector. The second term, the term with $x^1$, is the astigmatic focusing effect of the deflector and the third term, the term with $x^2$, is the second order aberration which is a comatic term[4]. This term deforms the spot shape into a comet and trefoil blur hence limiting the smallest discernible angle. For a homogeneous deflection field, as considered here, only the $\phi'(0)^3$ term contributes to the $x^2$ aberrations and the effect of other terms vanish. This is not however, necessarily the case for other monochromator designs. For instance, in the UC where the beam is traversing off-axis through a lens with spherical aberrations, $\phi'''(0) \neq 0$. There has been a recent attempt to mitigate this issue by using an off axis micro-lens[8].

The potential of a homogenous deflector can be written as

$$\phi = \phi_0 + \frac{\Delta\phi}{D} x \qquad 8$$

Where $\Delta\phi$ is the potential difference across an ideal deflector with a separation distance $D$ between the electrodes. Using Equation 8 in Equation 7 and integrating it over a deflector length $L$ yields

$$\theta = \frac{L\Delta\phi}{2D\phi_0} - \frac{xL}{4D^2}\left(\frac{\Delta\phi}{\phi_0}\right)^2 + \frac{x^2 L}{D^3}\left(\frac{3\Delta\phi^3}{16\phi_0^3}\right) \qquad 9$$

---

[2] Equation 5 resembles the fundamental theorem of calculus in 3 dimensions: $\Delta\vec{v} = \int \vec{\nabla} v \, ds = \eta \int \vec{\nabla} \phi^{\frac{1}{2}} \, ds$. Whereas a conventional line integral is valid for any trajectory, this integral is only an approximation when $ds$ does not follow the actual trajectory of the particle.

[3] Differs with a factor 2 from $\eta$ in [6]

[4] We might actually be able to correct higher order aberrations, leaving no intrinsic resolution limit due to geometric aberrations. However, a finite number of correctors will lead to a finite termination of this series expansion. Moreover, we assume here that we want to see how far we can push the resolution of monochromators without the inclusion of such correctors that complicate the design.

$$= \theta_0 - \left(\frac{x}{L}\right)\theta_0^2 + \frac{3}{2}\left(\frac{x}{L}\right)^2 \theta_0^3 \qquad 10$$

where $\theta_0 = \frac{L\Delta\phi}{2D\phi_0}$. Similar to $d_s$ deteriorating the smallest discernible radius $d_p$ in a focussed probe according to Equation 1, the angular comatic term here leads to a deterioration of angular resolution $d\alpha_{Co}$ of

$$d\alpha_{Co} \approx \frac{3}{2}\left(\frac{r}{L}\right)^2 \theta_0^3, \qquad 11$$

at the outer edges of a round beam. Here, the $x$-coordinate is equal to the beam radius $r$.

## 2.2 Power supply instability

Inverse time-of-flight of electrons in a monochromator is much higher than the bandwidth of any electrostatic power supply connected to the electrodes. For instance, even in a LVSEM at $\phi_0 \leq 1$ kV, the electrons traverse the column at $2\times 10^7$ m/s. This means that for a meter of column length, any interference slower than $2 \times 10^7$ Hz can be considered static drift. Even though small drifts can cause a dispersed beam to shift with respect to the energy selection slit, the instantaneous energy resolution for any wavepacket entering a monochromator is not subject to any time-dependence of the electric field. However, this does cause the nominal passing energy of the monochromator to drift. In EELS, one does not necessarily suffer from these drifts, since one can link together the deflector supplies of both the monochromator and analyser to symmetrically compensate for this nominal energy drift [9].

For LVSEM imaging purposes though, the time-averaged energy resolution integrated over the acquisition time per image is a measure of the practical energy resolution. Monochromators typically reach energy resolutions close to the theoretical specifications for only a ms or less, while relying on power supplies with a relative accuracy of several parts per billion[10]. The sensitivity monochromator deflectors due to voltage fluctuations affecting $\theta$ can be found by expanding around $\Delta\phi$ to yield the angular spread due to voltage fluctuations $d\alpha_s$ as

$$d\alpha_s = \frac{L}{2D\phi_0}\delta\phi = \theta_0 \frac{\delta\phi}{\Delta\phi} \qquad 12$$

From this intermediate result, we can already conclude that for a $d\phi$ minimally dependent on $\Delta\phi$ due to a small $d\alpha_s$, $\Delta\phi$ should be as large as possible to minimize drifts in the deflection angle. This is achieved by decreasing $L/D$ as much as possible.

## 2.3 Misalignments

Misalignments of deflector electrodes result in parasitic aberrations degrading the energy resolution in the absence of additional correction elements. Here, the effect of an antiparallel rotation of the two deflector plates relative to each other around their midpoints in the rotational $z$-direction, as depicted in Figure 4 is presented. If deflector plates rotate with an angle $\varphi/2$ in opposite directions around their midpoint at $y = 0$, the resulting distance $D$ between the electrodes is

$$D(y) = D_0 - 2y \tan\frac{\varphi}{2} \approx D_0\left(1 - \frac{y}{D_0}\varphi\right) \qquad 13$$

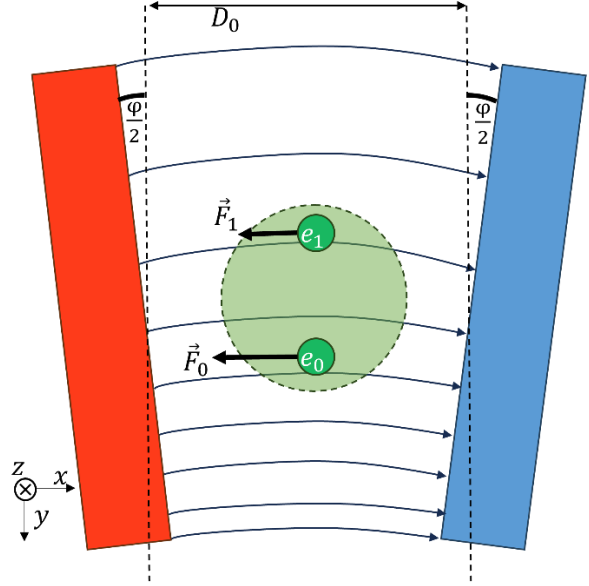

Figure 4: The deflector plates (red and blue) are both rotated anti-parallel with a small angle $\frac{\varphi}{2} \ll 1$. This results in an inhomogeneous deflection force experienced by different parts of the beam (indicated by example electrons $e_0$ and $e_1$).

Substituting this linear approximation in our expression for $\theta_0$ results in a contribution $d\alpha_\varphi$

$$d\alpha_\varphi := \theta_0 \frac{y}{D_0}\varphi \approx \theta_0 \frac{r}{D_0}\varphi. \qquad 14$$

The loss of angular resolution is due to rays traversing the deflector at the edge of the round beam, where $y \approx r$. The linear defects will dominate first (we can make $\varphi$ arbitrarily small to make this true) and higher order aberrations will involve more complicated correction elements than the linear aberrations. Therefore, we will base the mechanical tolerances for a given energy resolution on this linear effect only.

## 2.4 Coulomb interactions

From Equation 2 we could infer that $\phi_0$ can arbitrarily be lowered to improve the energy resolution. Though this is geometrically the case, other effects will then become dominant. Assuming pencil beam energy broadening[5], the Boersch effect directly adds to the energy spread of the beam as[11]

$$d\phi_B = K \frac{m}{\epsilon_0 e^2} \frac{I^2 L}{\phi_0}. \qquad 15$$

where $\varepsilon_0$ is the vacuum permittivity and dimensionless constant $K = 0.642$ is a constant that depends on initial statistical distributions and cutoff criteria. Including the Boersch effect with any inverse power $\frac{1}{\phi_0^n}$ leads to an optimal $\phi_0$ and thus a best theoretical energy resolution from first principles. However, the broadening in Equation 15 describes an approximation that is made for cylindrically symmetric optics. For a monochromator, the statistical energy broadening before the deflection fields that sort the electrons based on energy is irrelevant for the energy resolution of the system. Therefore, the relationship in Equation 15 should be taken with caution for monochromators. The topic is quite rich in detail for dispersive optics, and some configurations are claimed to show an "inverse Boersch effect"[12], lowering $d\phi$. Thus, more modelling for different types of dispersive optics is required to quantify the Boersch broadening in specific monochromator designs. However, Equation 15 does provide a sense of the magnitude and scaling of the statistical broadening when lowering $\phi_0$ to improve $d\phi$ and is thus used here. Instead of contributing to angular resolution loss, this directly adds to the energy resolution loss in a monochromator. In addition, statistical coulomb effects in a monochromator cause trajectory displacements. In the pencil beam regime, the angular resolution loss is given by [11]

$$d\alpha_T = \frac{C_p\, m^{\frac{3}{2}}\, I^3 L r}{\epsilon_0 e^{\frac{7}{2}}\, \phi^{\frac{5}{2}}} \qquad 16$$

where numerical constant $C_p = 8.31 \times 10^{-4}$. Again, there is an inverse dependence on $\phi$.

---
[5]Specifically, the pencil beam regime is picked since it's valid for the lowest energy broadening regime and it does not require additional analysis of the shape of the beam.

## 3 Practical energy resolution

In previous section, the effect of different contributions on the angular dispersion of an electrostatic deflector is analysed. The smallest discernible angle $\alpha_{\text{tot}}$ results from the linear combination[6]

$$\alpha_{\text{tot}} = \alpha + d\alpha_{Co} + d\alpha_s + d\alpha_\varphi + d\alpha_T \qquad 17$$

which includes all sources of angular broadening. Here, $\alpha$ is the current carrying angle whereas the other contributions merely blur this angle. Equating $\alpha_{\text{tot}}$ to $d\theta$ in Equation 2, the corresponding $d\phi$ is calculated. Moreover, adding the the Boersch effect to $d\phi$ gives $d\phi_{\text{tot}}$, as

$$d\phi_{tot} = d\phi_\alpha + d\phi_{Co} + d\phi_s + d\phi_\varphi + d\phi_T + d\phi_B \qquad 18$$

$$= \frac{\phi}{\theta}\alpha + \frac{3}{2}\frac{I}{B_r \pi^2}\left(\frac{\theta}{\alpha L}\right)^2 + \frac{\phi}{\Delta\phi}\delta\phi + \sqrt{\frac{I\phi}{B_r\pi^2}\frac{\varphi}{D\alpha}} + \frac{C_p\, m^{\frac{3}{2}}}{\epsilon_0 e^{\frac{7}{2}}\pi}\frac{I^{\frac{7}{2}}L}{\theta\phi^2 B_r^{\frac{1}{2}}\,\alpha} + \frac{Km}{\epsilon_0 e^2}\frac{I^2 L}{\phi}. \qquad 19$$

Here, the radius of a round beam is replaced with $r = \sqrt{\frac{I}{B_r\pi^2\alpha^2\phi}}$, $I$ is the unfiltered current and $B_r$ is the reduced brightness. Notice the indices of $\theta_0$, $\phi_0$ and $D_0$ and the negative sign of $d\phi$ have been dropped. By differentiating this equation w.r.t $\alpha$, an optimum $\alpha$ can be found. However, this does not necessarily lead to a better insight about different contributions unless they are compared one by one. First we assume that the energy resolution is dominated by geometric angle and contribution from coma. The optimal beam angle is found by equating $d\phi_{Co} = d\phi_\alpha$ as

$$\alpha = \left(\frac{3I}{2B_r\pi^2\phi L^2}\right)^{\frac{1}{3}}\theta. \qquad 20$$

Substituting this $\alpha$ in $d\phi_\alpha$ gives the optimum resolution in the presence of geometric and coma contributions.

---

[6] Whether linear, RMS or any other addition is chosen depends on assumptions concerning statistics. Though the answers are all different in their value, the procedures are mathematically equivalent and depend on the initial statistical distributions which we have no general information of. Thus, for simplicity, linear addition is chosen here.

$$d\phi_{\alpha,Co} \coloneqq 2d\phi_{Co} = 2d\phi_\alpha = \left(\frac{12I\phi^2}{B_r\pi^2 L^2}\right)^{\frac{1}{3}}. \qquad 21$$

Solving $d\phi_{\alpha,Co} = d\phi_{B,p}$ leads to the optimum potential

$$\phi = IL\left(\frac{B_r\pi^2}{12}\right)^{\frac{1}{5}}\left(\frac{Km}{\epsilon_0 e^2}\right)^{\frac{3}{5}}. \qquad 22$$

Substituting Equations 20 and 22 in Equation 21 leads to a new expression $d\phi_{\text{tot}}$

$$d\phi_{\text{tot}} = d\phi_{\alpha,Co,B} + d\phi_s + d\phi_\varphi + d\phi_T \qquad 23$$

$$= 2I\left[\left(\frac{Km}{\epsilon_0 e^2}\right)^2\left(\frac{12}{B_r\pi^2}\right)\right]^{\frac{1}{5}} + \frac{\phi}{\Delta\phi}\delta\phi + \frac{\varphi L^{\frac{3}{2}}I}{D\theta 3^{\frac{1}{2}}}\left(\frac{Km}{\epsilon_0 e^2}\right)^{\frac{1}{2}} + \frac{2C_p\, m^{\frac{1}{2}}I^{\frac{3}{2}}}{\pi B_r^{\frac{3}{6}}e^{\frac{3}{2}}K\theta^2}, \qquad 24$$

The first term in this equation,

$$d\phi_{\alpha,Co,B} = 2I\left[\left(\frac{Km}{\epsilon_0 e^2}\right)^2\left(\frac{12}{B_r\pi^2}\right)\right]^{\frac{1}{5}}, \qquad 25$$

which is a combined effect from $d\phi_\alpha$, $d\phi_{Co}$ and $d\phi_B$, presents the intrinsic resolution for an ideal monochromator free from engineering challenges namely mechanical imperfections and power supply instabilities. The only parameters influencing this equation are those imposed by "nature". Although trajectory displacement is also an intrinsic effect, it is excluded from this equation as it will shown its effect can be kept to a negligibly small value.

It should be noted that in the calculation of $d\phi_{\alpha,Co,B}$ it is assumed that $d\phi_{Co} = d\phi_\alpha = \frac{d\phi_B}{2}$. This assumption counts for a better energy resolution to the cost of a considerable brightness loss due to a larger contribution from coma.

For a diffraction limited beam the current $I$ is replaced with coherent current in the source as $I = 10^{-18}B_r$ [5] showing that the optimum resolution only depend on the reduced brightness of the source. For example, for a Schottky source with a $B_r = 10^8 \text{Am}^{-2}\text{V}^{-1}\text{sr}^{-1}$, and a coherent current of 100 pA, the diffraction limited resolution is 0.1 mV. For LVSEM applications, where often a higher current of $I = 10$ nA is preferred, this leads to $d\phi_{\alpha,Co,Bp} = 12$ mV.

In practice, however, this level of resolution isn't easily achievable. This is attributed to the three remaining terms in Equation 24. A requirement is derived from each term. First, trajectory displacement is considered. By setting $d\phi_T \ll d\phi_{\alpha,Co,B}$, and rearranging the terms

$$\theta \gg \sqrt{\frac{C_p\, m^{\frac{1}{10}}I^{\frac{1}{2}}\epsilon_0^{\frac{2}{5}}}{12^{\frac{1}{5}}\pi^{\frac{3}{5}}B_r^{\frac{3}{10}}e^{\frac{7}{10}}K^{\frac{7}{5}}}} = I^{\frac{1}{4}}\left(\frac{C_p^{10}m\epsilon_0^4}{B_r^3 12^2 \pi^6 e^7 K^{14}}\right)^{\frac{1}{20}} \approx 16\frac{I^{\frac{1}{4}}}{B_r^{\frac{3}{20}}}. \qquad 26$$

Here, the constants of nature and statistical constants can be rounded to a value of 16 in standard units. For $I = 10$ nA, $B_r = 10^8$ Am$^{-2}$rad$^{-2}$V$^{-1}$, $\theta \gg 10$ mrad. In the diffraction limit, where using $I = 10^{-18}B_r$, $\theta \gg 5 \times 10^{-4}B_r^{\frac{1}{10}} = 3$ mrad for $B_r = 10^8$ Am$^{-2}$rad$^{-2}$V$^{-1}$.

Moreover, it so happens that for $B_r = 10^8$ Am$^{-2}$rad$^{-2}$V$^{-1}$, $B_r^{\frac{3}{20}} = 16$. Assuming sources in new SEM's to have a brightness within a factor of 10 of this value, the result for $B_r^{\frac{3}{20}}$ will not vary more than a factor $10^{\frac{3}{20}} \approx 1.4$, which we assume to be negligible for typical scaling purposes. Therefore, one might typically assume the requirement $\theta \gg I^{\frac{1}{4}}$ in standard units using the Pencil beam approximation. Though $\theta \gg 10$ mrad could lead to some engineering challenges, this is not an intrinsic problem with monochromators.

In order to explore what really limits practical monochromator performance, we will have a look at the tolerances. First, we examine the deflection potential tolerances. Now, we will get the relative tolerance of the power supply by demanding $d\phi_s \ll d\phi_{\alpha,\text{Co,B}}$. As an example, we assume $L = 0.1$ m, which yields

$$\frac{\delta\phi}{\Delta\phi} \ll \frac{2I}{\phi}\left[\left(\frac{Km}{\epsilon_0 e^2}\right)^2\left(\frac{12}{B_r\pi^2}\right)\right]^{\frac{1}{5}} = \frac{2}{L}\left(\frac{12}{B_r\pi^2}\right)^{\frac{2}{5}}\left(\frac{\epsilon_0 e^2}{Km}\right)^{\frac{1}{5}} \approx 3\times 10^{-6}. \qquad 27$$

This is still possible for LVSEM with a stable power supply ($\gg$ 18.4 bits resolution). For TEM though, $\delta\phi$ is on top of an acceleration potential. For example[7], $\delta\phi$ for $D = 1$ mm and $\theta = 1$ rad gives

$$\delta\phi \ll \frac{\Delta\phi}{\phi}d\phi_{\alpha,\text{Co,B}} = \frac{4I\theta D}{L}\left[\left(\frac{Km}{\epsilon_0 e^2}\right)^2\left(\frac{12}{B_r\pi^2}\right)\right]^{\frac{1}{5}} \approx 2\times 10^{-4}\text{ V}. \qquad 28$$

Though maintaining this accuracy by itself is possible, stacking the deflection voltage supply on top of an acceleration potential of 100 kV demands a relative stability of $\ll 2\times 10^{-9}$. This does pose a more serious engineering demand, and explains the limited stability time that can be in the ms range [13] [10].

In order to compute mechanical alignment tolerances, we require $d\phi_\varphi \ll d\phi_{\alpha,\text{Co,B}}$ which yields

$$\varphi \ll \frac{D\theta}{L^{\frac{3}{2}}}\left[\frac{12^7\epsilon_0 e^2}{B_r^2\pi^4 Km}\right]^{\frac{1}{10}} \approx 4\times 10^{-5}\text{ rad} \qquad 29$$

for the same illustrative values. In practice, this extremely low tolerance implies the incorporation of additional stigmators and other corrective elements to remove aberrations induced my misalignments. Thus, in designing a new type of monochromator, the tolerances of the electrostatic and mechanical components should be a main focus.

# 4 Simple electrostatic monochromator approach

Monochromators such as the ones mentioned in section 1 inherently suffer from the problems described in section 2 and 3. In order to circumvent these issues, monochromators rely on additional complex correction elements and electronics. In this section, a new approach towards a simple monochromator design is presented. Compared with conventional high performance monochromators, the complexity of monochromators is thereby considerably reduced while the performance is unaffected.

## 4.1 Electrostatic fringe field deflectors

Equation 12 and Equation 14 show that the electro-mechanical tolerances scale inversely with $D$. Therefore, this can be tuned for a better design. For an electrostatic deflector, if $D \approx L$, the fringe field becomes important and needs to be included in the calculation of $\theta$. When $D \gg L$, the fringe field contributions will dominate the shape of the potential, hence the name "fringe field deflector". In order to model the fringe field effect, its potential is approximated by two line charges, adapted from [14]

---

[7] 1 rad is usually not a "small" angle, but for $\theta \approx 2\sin\left(\frac{\theta}{2}\right)$, but the difference is <5%. Though this difference requires consideration for implementation in a design, this does not fundamentally change the first order scaling. and we will assume the aberrations scale the same for these large angles. Progressing to large angles with any method would add contributions, while still including the effects mentioned so far.

$$\phi = \phi_0 + \frac{\Delta\phi}{4} \ln \frac{\sqrt{\left(x + \frac{D}{2}\right)^2 + z^2}}{\sqrt{\left(x - \frac{D}{2}\right)^2 + z^2}} \qquad 30$$

$$= \phi_0 + \frac{\Delta\phi}{8} \ln \frac{\left(x + \frac{D}{2}\right)^2 + z^2}{\left(x - \frac{D}{2}\right)^2 + z^2}, \qquad 31$$

Where $\Delta\phi$ and $D$ are chosen such that the electric field strength at the centre of the deflector the same[8] as for a homogeneous deflector. Now, applying the same mathematical procedure as in subsection 2.1 requires us to calculate the higher order derivatives around $\phi(x = 0)$. Taking the higher order derivatives of Equation 30 to $x$ and filling in $x = 0$, results in

$$\phi(0) = \phi_0 \qquad 32$$

$$\phi'(0) = \Delta\phi \frac{D}{D^2 + 4z^2} \qquad 33$$

$$\phi''(0) = 0 \qquad 34$$

$$\phi'''(0) = \Delta\phi \frac{8D(D^2 - 12z^2)}{(D^2 + 4z^2)^3} \qquad 35$$

Substituting these derivatives in Equation 7, and integrating over $z$ from $-\infty$ to $\infty$ yields

$$\theta_f = \theta_{f,0} - \frac{x}{D} \frac{\theta_{f,0}^2}{\pi} + \left(\frac{x}{D}\right)^2 \frac{9\theta_{f,0}^3}{4\pi^2} \qquad 36$$

where the nominal deflection angle $\theta_{f,0} = \frac{\pi\Delta\phi}{4\phi_0}$, which is no longer dependent on $D$, effectively eliminating the first-order misalignment problems contributing to $d\phi_\varphi$.

The contribution $d\phi_s$ to Equation 24 remains unchanged when transitioning from homogeneous to fringe deflection fields. However, $d\phi_s = \frac{\phi}{\Delta\phi}\delta\phi = \frac{L}{2D\theta}\delta\phi$ for homogeneous deflection fields, compared to $d\phi_s = \frac{\pi}{4\theta}\delta\phi$ for fringe fields.

This implies a difference in sensitivity to $\delta\phi$, with a ratio of $\frac{2L}{\pi D}$. Therefore, a deflector dominated by a homogeneous field, particularly one where the deflector length $L$ is much greater than the gap $D$, is more sensitive to $\delta\phi$, which is generally undesirable.

In contrast, fringe-field-dominated configurations can ideally achieve $d\phi_s \approx \delta\phi$, resulting in minimal sensitivity to angular deviations. This is preferable, provided that the required energy dispersion $\Delta\phi$ can still be achieved without triggering a plasma discharge (avoiding electrical breakdown).

Although further improvements might be attempted, in practice the drift of the acceleration potential in a TEM becomes the dominant source of error, since it typically scales as $d\phi_s \approx \delta\phi$. In the case of LVSEM, this particular limitation may not apply. However, in practical implementations, there is often still a residual contribution to $d\phi_s$, that scales as $d\phi_s \geq \delta\phi$.

For the potential described in [14] for finite cylinders, the parameter $\frac{D}{2}$ in our result Equation 36 is replaced by $\sqrt{d^2 - a^2}$, where $d$ is half the distance between the centres of the cylinders and $a$ is the radius of the cylinders. Therefore, this does not alter any of the conclusions made about line charges so far. However, the factor $\Delta\phi/4$ is replaced by $\frac{1}{2}\Delta\phi g$, where

$$g := \frac{1}{\ln(\hat{d}+\sqrt{\hat{d}^2-1})}, \quad\quad 37$$

and $\hat{d} := d/a$. This does introduce a new sensitivity to rotational misalignments in the final result. Analysing the relative sensitivity of $g$, we have

$$\frac{\partial g}{\partial \hat{d}}\frac{\hat{d}}{g} = \frac{\hat{d}}{\sqrt{\hat{d}^2-1}\ln(\hat{d}+\sqrt{\hat{d}^2-1})}. \quad\quad 38$$

---

[8] Any other amplitude results in a different $\theta_{f,0}$, but the scaling as a function of $\theta_{f,0}$ remains the same.

Though this yields a finite tilt sensitivity, it does converge to 0 for $\hat{d} \gg 1$, where we approach the line charge model.

## 4.2  Beam energy variation

As discussed in subsection 4.1, we can neglect $d\phi_\varphi$ for fringe fields. We assume $\Delta\phi \approx \phi_0$ to minimize $d\phi_s$ and hence $\theta_{f,0} \approx \frac{\pi}{4}$, leading to $d\phi_s \approx \delta\phi$. Finally, for any current applicable in an SEM, $\theta_{f,0} \approx \frac{\pi}{4} \gg I^{\frac{1}{4}}$ means we can neglect trajectory displacement. Therefore, the remaining contributions are

$$d\phi_{\text{tot}} = d\phi_{\alpha,\text{Co},B} + d\phi_s \quad\quad 39$$

$$= 2I\left[\left(\frac{Km}{\epsilon_0 e^2}\right)^2 \left(\frac{12}{B_r \pi^2}\right)\right]^{\frac{1}{5}} + \delta\phi \quad\quad 40$$

where the value of $d\phi_{\alpha,\text{Co},B}$ is left based on homogeneous fields. Though the results for $d\phi_{Co}$ differ by a factor $\frac{3/2}{9/4\pi^2}$ between the fringe field and homogeneous field when $D = L$, only the scale of the final result is considered, while the numerical factor of $d\phi_{Co}$ appears with a power $\frac{1}{5}$ in the final result and is thus relatively inconsequential. The goal is not to get an exact numerical pre-factor, but instead to derive a general scaling law. To get to this conclusion, we had to assume $L \approx D$, which fixes $\phi$ according to Equation 22. Assuming there is space for only $D = L = 2$ mm because there is a limited budget for vacuum pumps and because electrostatic fields interacting with vacuum chamber walls cannot be controlled with much larger shapes, the optimal $\phi = 85$ V is much lower than any typical gun energy. Moreover, increasing the size and beam energy would push the deflection voltages, which become difficult to feed into the microscope when passing the $\approx 1$ kV range without creating discharges. It is important to note that this is the optimal beam energy in the deflection space only. In order to minimize Boersch broadening as much as possible, we should therefore also limit this low beam energy to the deflection space only[9]. Though this is possible, in practice it means that the beginning and end of the deflection space will effectively both become lenses with short focal lengths. In order to minimize the geometric aberrations of such deceleration and acceleration lenses, they should be of similar scale in size. With a smaller size, the spherical aberrations of this lens will dominate the energy resolution, while with a bigger size, the deflectors themselves will dominate the deceleration and acceleration lens fields, introducing large astigmatism that also needs additional correction elements. Thus, an optimal balance of all factors contributing to $d\phi$ may be achieved by designing a monochromator entirely based on fringe fields, where the fields of the dispersive deflector and transfer lenses overlap. One way to design, build and align such small electrodes on top of each other using microfabricated electrodes, through MEMS techniques. In such a design, the fringe fields of the deflector electrodes and deceleration and acceleration lenses will all have similar spatial dimensions, and should therefore be modelled collectively, while taking their interactive properties into account. Due to the typical voltages and electric fields achievable between such a MEMS based monochromator, the deflection field of the monochromator would have to be mounted before the acceleration potential in systems like (S)TEM's.

---

[9] This approach is partially utilized in [15], where the beam is decelerated before deflection. However, this configuration is sub-optimal because the beam drifts at the same energy before reaching the selection aperture, creating unnecessary drift space at low beam energy, increasing susceptibility to Boersch broadening and stray field interference.

# 5 Conclusion

Electrostatic or magnetostatic deflection fields are used in current monochromators as main dispersive elements. Using a thin deflector model, the intrinsic energy resolution of monochromators is shown to be independent of the dimensions of the system, when the beam parameters are properly scaled. This results in an energy resolution of 0.1 meV in the diffraction limit. For homogeneous deflection fields, it is shown that the deflector requires extreme mechanical accuracies ($\ll 4 \times 10^{-5}$ rad), which isn't easily achievable with current fabrication technologies. Therefore, for suppressing parasitic aberrations introduced during fabrication of the components, additional correcting elements are added to the design. Since the inclusion of more correction elements requires fine-tuning of more power supplies, the inclusion of these additional correction elements can deteriorate practical resolution limit and user-friendliness while increasing cost. The tight mechanical tolerance requirement can be circumvented to a great extent using a fringe-field dominated design, which simultaneously relaxes the requirements on power supply stability. The optimal beam energy resolution is achieved by decelerating the beam shortly before the main dispersive deflector, to suppress the Boersch effect. The compact nature of a fringe-field deflector and the sudden short-range decelerating and accelerating fields around it makes MEMS techniques for fabricating the components / electrodes favourable. To the best of our knowledge, such a fully electrostatic MEMS monochromator has not been realized yet. We acknowledge that companies might have derived scaling laws similar to the ones presented here, or even more applicable to a particular design. However, we have not been able to publicly find general design criteria as presented here.

# 6 Appendix

## 6.1 Parabolic trajectory

For a homogeneous deflection field, instead of assuming any trajectory for the electron in a deflector, since

$$\phi = \phi_0 + \frac{\Delta\phi}{D}x \qquad 41$$

we have $\frac{dv_{z,1}}{dt} = 0$ inside the deflection field. However, it is different from $v_{z,0}$ outside the deflector because of the acceleration at the boundaries of the deflector. Assuming a sharp cut-off between the $\phi = \phi_0$ outside the deflector and inside (Equation 41) the deflector, the electron experiences an instantaneous acceleration in the z-direction such that by conservation of energy, the z-velocity inside the deflector $v_{z,1}$ (through conservation of energy) is given by

$$\frac{1}{2}mv_1^2 = \frac{1}{2}m(v_{z,1}^2 + v_{x,1}^2) = e\phi \qquad 42$$

$$v_{z,1} = \sqrt{\frac{2}{m}}\sqrt{e\phi_0 + \frac{e\Delta\phi}{D}x - \frac{1}{2}mv_{x,1}^2}. \qquad 43$$

Assuming symmetric deflection, the x-velocity before and immediately after entering the deflector respectively $v_{x,0} = v_{x,1} = \frac{1}{2}\Delta v_x$. Then, a constant electrostatic force in the x-direction over a time interval $\Delta t = \frac{L}{v_{z,1}}$ with substitution of equation 43 yields

$$\Delta v_x = \frac{e\Delta\phi}{mD}\frac{L}{v_{z,1}} = \frac{e\Delta\phi L}{mD}\frac{1}{\sqrt{\frac{2}{m}}\sqrt{e\phi_0 + \frac{e\Delta\phi}{D}x - \frac{1}{8}m\Delta v_x^2}} \qquad 44$$

Then, by dividing by $v_0$ to get $\tilde{\theta} := \frac{\Delta v_x}{v_0} = 2\sin\frac{\theta}{2}$, where $v_0 = \sqrt{\frac{2e\phi_0}{m}}$ this simplifies to

$$\tilde{\theta} = \frac{\theta_0}{\sqrt{1 + 2\theta_0 x - \frac{1}{4}\tilde{\theta}^2}}, \qquad 45$$

where $x := \frac{x}{L}$ and again $\theta_0 := \frac{L\Delta\phi}{2D\phi_0}$. Squaring both sides and solving for $\tilde{\theta}^2$ we get

$$\tilde{\theta}^2 = 2(1 + 2\theta_0 x) \pm 2\sqrt{(1 + 2\theta_0 x)^2 - \theta_0^2}, \qquad 46$$

where subtraction of the two terms represent the physical result since then $\tilde{\theta}(\theta_0 = 0) = 0$.
Then, taking the square root and of this result we get

$$\tilde{\theta} = \sqrt{2(1 + 2\theta_0 x) - 2\sqrt{(1 + 2\theta_0 x)^2 - \theta_0^2}}, \qquad 47$$

where we can finally retrieve $\theta$ as

$$\theta = 2\arcsin\frac{1}{2}\tilde{\theta} = 2\arcsin\left(\frac{1}{2}\sqrt{2(1 + 2\theta_0 x) - 2\sqrt{(1 + 2\theta_0 x)^2 - \theta_0^2}}\right). \qquad 48$$

And then expanding the lowest order terms around $\theta_0 = 0, x = 0$ we retrieve

$$\theta \approx \theta_0 - x\theta_0^2 + \frac{3}{2}x^2\theta_0^3, \qquad 49$$

which is exactly the same as Equation 10. From this we can deduce that setting $ds = dz$ retrieves the exact result for small angle deflection.